**TITLE:**
California Community Colleges: Designing mentoring for access to social capital

**AUTHOR(S):**
Balaraman, A.[1]; Maokosy, S.[2,3]; Slaton, L.[4]; Cardona, R.[5]; Maokosy, P.[6]; Gaitan, N.[5]
**ORGANIZATION:**
[1] University of California-Berkeley; [2] Reedley College-California; [3] Porterville College-California; [4] Solano College-California; [5] Glendale College-California; [6] Coalinga College-California

## ABSTRACT

Successful careers are built on Skills (what you know), Occupational Identity (what you believe you can be), and Social Capital (who you know) (Fischer et al., 2018). Higher education spends significant resources to address the first, sometimes at the cost of the latter two, which are complex and expensive to promote. The NSF-supported research aimed to evaluate the effectiveness and adoption of a mentoring program and technology that is part of course instruction to broker social capital in California Community Colleges.

Social capital research has demonstrated the value of relationships and networks to enhance college opportunities for first-generation, underrepresented students (Beals et al., 2021). Mario Luis Small, a sociologist at Harvard University, contributed to the literature on social capital, the idea that institutions such as Institutions for Higher Education (IHE) can be brokers of social capital (Small, 2009). This research explores explicitly how near-peer mentoring programs, rather than stand-alone opt-in guidance, can be integrated into the instruction/pedagogy by faculty at California community colleges.

The research was conducted at 5 California community colleges (Reedley, Porterville, Coalinga, Solano, and Glendale) – predominantly serving first-generation, low-income, Hispanic-serving institutions. Knowing the crucial role faculty and classrooms play for students, especially in IHE (Inside Higher Ed. 2021), this research focused on measuring the effectiveness of mentoring programs (curriculum and technology) integrated with instruction by faculty across various courses. A mixed-methods approach was used to gather social cognitive measures of student self-efficacy, occupational identity, and social capital access (Vautero, Silva, 2022).

Measures were collected using survey instruments at the beginning and culmination of the mentoring program. One of the most consistent measures observed across all the pilots was increased student self-efficacy of skills and competencies (3% - 7%) across colleges, geographies, and course formats after <8-hour mentoring program. Additionally, the research offers insights into implementing peer and near-peer mentoring programs within various course pedagogies to improve course completion, with significant instructional and non-instructional (cost) advantages.

## LITERATURE REVIEW

Mentorship has long served a vital role in education and professional development. Higher education institutions, especially in STEMM (Science, Technology, Engineering, Math, and Medicine), have adopted mentorship as a powerful way to attract, develop, and retain a diverse workforce necessary to power scientific innovations and talent to be a global leader in STEMM. (NAS, 2019) Besides STEMM, mentorship is also an effective way to augment and accelerate guidance for a wide range of outcomes- college and career planning, support of marginalized youth, encourage college-going aspirations, improve grades and test scores, increase belonging and agency, among others (Trepanler-Street, 2007; Albright, 2017; Griffin, 2015). Despite the importance of mentoring, it is not a formalized part of education and training. With a few exceptions, it happens ad hoc through opt-in programs. A Gallup poll found that just 22% of science and engineering majors strongly felt they had a mentor during their undergraduate studies (Gallup, 2018). Further stratification in mentorship was evident, with students in 2-year colleges being significantly less likely to have mentors than their 4-year college counterparts and students in 4-year public colleges being less likely to have mentors than their 4-year private college counterparts (Inside Higher Ed, 2021).

Recent research by Raj Chetty and colleagues (Chetty, 2022) suggests a troubling prevalence of socio-economic status-based friendships in colleges and its correlation to economic outcomes- "Rich people make friends of college classmates, poorer people make friends of neighbors." Colleges play a significant role in teaching- developing skills and knowledge in a discipline. Concurrently, colleges also significantly expand the tacit and implicit knowledge necessary for their students' upward mobility through friendships, which are their social capital. There is an opportunity for colleges to develop economic connectedness across SES among their students to support post-secondary economic outcomes.

This study was aimed at investigating (a) the effectiveness of a mentoring program formalized as *part of instruction* in developing skills and knowledge in a discipline, (b) evaluating the role of mentoring in developing the ability to build networks, and social capital in a classroom (c) design near-peer mentoring programs to integrate into formalized instruction modalities at colleges.

## CONTENT

This research was funded by the National Science Foundation's Directorate of Translational Research and Technology Innovations Program of deep technologies—discoveries in fundamental science and engineering. Epixego's Inc.'s Artificial Intelligence (AI)- based deep technology was part of NSF's research project. The project's objective was to evaluate the technology's application to scale mentoring programs at five 2-year HSI (Hispanic Serving Institutions) colleges in California, predominantly serving first-generation, low-income students.

This project was conducted in two phases across the five colleges, working with faculty to iteratively design a near-peer mentoring program to accommodate variability in instructional environments across courses and faculty.

### Method

The project is designed to (a) leverage the critical role played by faculty in course instructional design and formalize a mentoring program as part of their course curriculum and (b) collaborate with faculty teaching discipline-agnostic coursework, e.g., entrepreneurship and innovation-related credit and noncredit Career Technical Education (CTE) courses. An outreach flier was sent to faculty members at California Community Colleges, and interested faculty members were interviewed and introduced to the project and research objectives.
The following hypotheses were tested in the colleges and courses participating in the project.

a) Designing mentoring programs that allow participants to translate their learning into competencies will help broaden participants' understanding of skills and competencies.
[Skills and Competency Measures]

b) Providing access to vicarious learning via near-peer role models expands occupational identities through vicarious learning.
[Occupational Identity Measures]

c) Providing robust curriculum and learning objectives, in addition to each participant being a mentor AND mentee, provides examples of a successful mentor-mentee relationship. Participants are likely to value the mentoring relationship in building their social capital.
[Social Capital Measures]

**Table 1: Summary of colleges, faculty, and courses where mentoring was incorporated into instruction**

| College Name | Role | Course/Program | # of Pilots/Phases [Summer, Fall, 2023] |
|---|---|---|---|
| Reedley College [2-year] | Adjunct Faculty | Operation of Small Business; Introduction to Business; Investment Fundamentals | Two pilots – Summer & Fall |
| Porterville College [2-year] | Adjunct Faculty | Quantitative Methods for Business Decision Making; Business Statistics | Two pilots – Summer & Fall |
| Solano College [2-year] | Tenured Faculty | Advisor, Entrepreneur Club, Introduction to Small Business and Entrepreneurship | One pilot – Fall |
| Glendale College [2-year] | Adjunct Faculty | Introduction to Business Communication Marketing and Communication; Introduction to Advertising | Two pilots – Summer & Fall |
| West Hills College [2-year] | Adjunct Faculty | Introduction to Ag-Business – CTE course | Two pilots – Fall |

**Data Collection and Analysis**

A mixed-methods approach of student surveys and semi-structured interviews was used to evaluate the effectiveness of the mentoring. The mentoring curriculum was incorporated over a faculty-determined 5-week period. The curriculum was embedded as part of the pedagogy for extra credit in the course instruction and assigned credit. Instructors invited their students to participate in the mentoring program via Zoom as part of course completion—either with mandatory participation, extra credit, or a combination thereof.

The mentoring program differed from traditional in-person mentoring programs in the following ways:

- Mentor recruitment and incentives: Peer and near-peers from the course were matched based on near-peer mentor preferences, obviating the need for external mentor recruitment. Additionally, each student was assigned a mentor and mentee to a near-peer, incentivizing them to participate and develop their skills.
- Mentor: Mentee Match: Technology (Epixego Inc.) facilitated near-peer suggestions, student mentor preferences, and matching. Each student was assigned a mentor and a mentee to engage in conversations via Zoom breakout rooms and/or individual phone/Zoom conversations.

Below is an overview of the 5-week mentoring curriculum with instruction, self-learning, and reflection components.

**Table 2: Mentoring curriculum at-a-glance**

| Week | Topic |
|---|---|
| Week 1<br>90 minutes | Orientation: Why 'co-create' your mentoring experience?<br>- Pre-program survey |
| Week 2<br>90 minutes | Dynamics of Mentoring Relationships<br>- Mentoring conversation in breakout rooms |
| Week 3<br>90 minutes | Challenges in Mentoring Relationships<br>- Mentoring conversation in breakout rooms |
| Week 4<br>90 minutes | Changing Occupational Identities & Social Capital<br>- Mentoring conversation in breakout rooms |
| Week 5<br>90 minutes | Cultivating a robust constellation of near-peer mentors<br>- Mentoring conversation in breakout rooms<br>- post-program survey + reflections |

Quantitative data was gathered from the student participants through survey instruments using a 5-point Likert scale at the beginning and end of the program. Qualitative data were obtained using open-text field entries and semi-structured interviews. Quantitative measures were adapted from the "Science of Effective Mentoring."; "Social Cognitive Career Theory," "Social exchange theory," "Social Capital Theory," and "Ecological Systems Theory" (NAS, 2019). Below is a sample of the survey measures collected.

**Table 3: Sample survey measures**

| Category | # of questions | Sample Questions |
|---|---|---|
| Self-efficacy: skills and competency | 5 | "I feel confident about my skills and competent in areas where I can apply them."<br><br>"I understand how to develop and translate my skills into the language of higher education and the labor market." |
| Access to social capital | 5 | "The people in my network have information and relationships that can help me in becoming a professional in my chosen field."<br><br>"I feel comfortable discussing my career interests and aspirations with other adults in my community." |
| Occupational Identity/ies | 5 | "I am clear about what I want to do professionally five years from now."<br><br>"I am aware of possible career paths available to me and feel confident about the resources that can guide me." |

**Results and Discussion**

The data (qualitative and quantitative data) indicate the benefits of incorporating a mentoring curriculum as part of a formalized instructional design. In this project, an overwhelming majority (90-95%) of the student participants reported not having mentors or being in mentorship programs, confirming earlier research that non-traditional learners in community colleges lack access to mentors or mentoring programs.

Table 4: Aggregated Mentoring Statistics

| College | Key Statistics | | | Avg. Increase in Pre vs. Post-Response [average computed across all responses in that category] | | |
|---|---|---|---|---|---|---|
| | # of participants | % of participants who never had mentors or been in mentoring programs | Mentoring program completion rate | SE (Skills & Competency) | Social Capital | Occupational Identity |
| Reedley + Porterville [RC + PC] | 100+ | 99.9% | 100% | 3.82% | 0% | 5.21% |
| Solano | 120+ | 90% | 70+% | 3.76% | 4.65% | 8.33% |
| Glendale | 50+ | 92% | 80+% | 4.27% | 7.14% | 7.01% |
| West Hills | 40+ | 60% | 90+% | 3.20% | 1.81% | 1% |

Over a 5-week, 8-hour program, the data indicate a consistent increase in student self-efficacy as related to their skills and competencies gained. Self-efficacy measures support the viability of a mentoring curriculum solution that can be offered as part of the learning and taught with a pedagogy similar to how skills are taught. The following conclusions can be drawn from the data analysis:

- Consistent increase in mentor and mentee self-efficacy across programs and formats. Self-metrics for skills and competencies measures, while appearing modest, reflect a relevant measure. The improvement in self-efficacy is within the anticipated range across similar studies measuring near-peer mentoring effectiveness [range between 0.25 - 3.0 units]. (Sun, Clark-Midura, 2022; Trujillo el al. 2015; Deane et al. 2022)

- Access to social capital measures was variable across the colleges. While the average aggregate social capital measures did not show change at Reedley and Porterville colleges (two colleges where the instruction and embedded mentoring programs were conducted asynchronously), the differences in responses indicate a more nuanced measurement necessary to distinguish the types of social capital. For example, There was a 5.8% increase post vs. pre for the social capital question: "*The people in my network have information and relationships that can help me in becoming a professional in my chosen field.*"; but no significant change was observed for another social capital-related question, "*The people in my network can speak to the quality of my accomplishments, attitudes, and work ethics.*"
- Occupational identity measures assess " the perception of one's occupational interests, capabilities, goals, and values." The mentoring program saw increased occupational identity facilitated by mentoring across the colleges and courses.

  > "*As a field worker at a farm, aspiring to be self-employed, I barely knew Taylor, who is in my class. I selected Taylor to be my mentor – like me, she is a promise program student, who enjoyed writing, but unlike me, she was self-employed in Sports Promotion at Salt Lake Bees. I was surprised to find that marketing and promotion could be a path to self-employment. We also collaborated on homework assignments.*" - Student, Reedley Community College.

  > "*I am studying Business Management at my community college and also working as a teacher aide and caregiver. I do aspire to be an entrepreneur and am taking courses to enable me to have my own business, perhaps a childcare business. Sharon was a mentor suggested to me, who is not only in the same program, who I did not know, but we share an interest in digital design courses. Sharon runs her own digital advertising and marketing firm, and I now think there may be more entrepreneurial options for me besides child care.*" - Student, Solano Community College.

*Effectiveness of designing instruction and near-peer mentoring*: Our data confirms earlier research (NAS, 2019)- higher education has formalized education and training. However, functions and behaviors that support learning, such as mentoring, have been mainly unavailable to students, especially those who need it the most. This study presents a significant opportunity to enhance mentorship processes for the 'unsavvy' those who have never had mentors in the past. Incorporating peer/near-peer mentoring in the classroom, as part of pedagogy, allows the instructors to implement-

a) strategies to allow students to navigate myriad course-related challenges with the psychosocial and academic support of their near-peers, without power barriers, and
b) opportunities for students to develop learning strategies from peers and near-peer mentors while consolidating their learning and knowledge as mentors.

Making mentoring available to the 'unsavvy' happens most effectively in the classroom. As Anthony Abraham Jack, a junior fellow at the Harvard Society of Fellows and an assistant professor at the Harvard Graduate School of Education, notes, "Students from a low-income household… entering college from a local distressed public high school, may well have been told

'just keep your head down and do good work.' To these students, mentorship seems like the 'wrong way to get ahead…They are more tasked with maintaining order than making connections."

The opportunity to enhance the mentoring process may lie in thinking of social capital education, which is much like skills. Becoming 'savvy' with mentoring and growing social capital effectively supports students in learning discipline-specific, professional, and cultural tacit skills to help them navigate toward becoming successful professionals in their post-secondary lives. IHEs can re-frame the budgeting of mentoring program models from 'support services' to 'core instruction' budgets at a significant cost advantage (average cost of $2,500/student to $10/student).

**Conclusion**

Social capital is a significant factor in college and post-college career success. However, many first-generation, low-income, and underrepresented students have limited access to these networks, which has driven racial and gender wealth gaps. Mentoring and experiential learning are the most effective high-impact practices to level the playing field for information, opportunity, and support to social networks and career-connected learning. The case for mentoring interventions that are included as part of instruction to strengthen social capital, rather than opt-in programs is compelling- First-generation students are less likely to engage in internships and social capital–building activities, according to a report from Strada's Center for Education Consumer Insights (Strada, 2023).

While educational technologies have long focused on providing tools for educational content, assessment, and productivity, their efforts to provide tools for enhancing social capital are a relatively new phenomenon. Today, there is ample research that indicates the presence of institutional power barriers, such as in faculty mentor programs, and lack of deep-level shared values, such as shared attitudes, goals, interests, values, learning styles, and competencies in mentoring programs based on demographic factors alone; and the ineffectiveness of shared values by matching mentors and mentees- based on narrowly stated goals of the mentee, and matching them to based on a mentor's biography. Notably, programs designed for first-generation, low-income students to engage in internships and social capital-building activities such as mentoring do not have instructional content or support. As a result, only savvy students get guidance, and there is a racial and gender pattern among the savvy students. "...History has borne out the reality that mentoring programs intended to serve women disproportionately benefit white women, and programs intended to serve minorities mainly benefit minority males" (Ong et al., 2018). This study amongst California Community Colleges provides a playbook for implementing mentoring into a core aspect of instruction to allow teaching and helping their students develop social capital, regardless of the course format.